\documentclass[preprints,article,accept,moreauthors,pdftex]{mdpi}
\usepackage{longtable}
\usepackage{booktabs}
\usepackage{changes}
\definechangesauthor[name={Per cusse}, color=orange]{per}
\newcommand{\physrep}{{Phys. Rep.}}

\newcommand{\mnras}{{Mon. Not. R. Astron. Soc.}}
\newcommand{\apj}{{Astrophys. J.}}

\newcommand{\aap}{{Astron. Astrophys.}}
\newcommand{\apjs}{{Astrophys. J. Suppl. Ser.}}

\newcommand{\apjl}{{Astrophys. J.}}
\newcommand{\pasj}{{Publ. Astron. Soc. Jpn.}}
\newcommand{\procspie}{{Proc. SPIE Int. Soc. Opt. Eng.}}

\newcommand{\pra}{{Phys. Rev. A}}
\newcommand{\jcp}{{J. Chem. Phys.}}
\newcommand{\prl}{{Phys. Rev. Lett.}}

\firstpage{1} 
\pubvolume{1}
\issuenum{1}
\articlenumber{0}
\pubyear{2021}
\copyrightyear{2020}
\datereceived{} 
\dateaccepted{} 
\datepublished{} 
\hreflink{https://doi.org/} % If needed use \linebreak
\Title{Uncertainties in Atomic Data for Modeling Astrophysical Charge Exchange Plasmas}

\TitleCitation{Uncertainties in Atomic Data for Modeling Astrophysical Charge Exchange Plasmas}

\Author{Liyi Gu $^{1,2 \dagger}$\orcidA{} and Chintan Shah $^{3,4,5}$\orcidB{} and Ruitian Zhang $^{6,7}$}

\AuthorNames{Liyi Gu, Chintan Shah, and Ruitian Zhang}

\AuthorCitation{Gu, L., Shah, C. and Zhang, R.}
\address{%
$^{1}$ \quad SRON Netherlands Institute for Space Research, Niels Bohrweg 4, 2333 CA Leiden, the Netherlands; l.gu@sron.nl\\
$^{2}$ \quad RIKEN High Energy Astrophysics Laboratory, 2-1 Hirosawa, Wako, Saitama 351-0198, Japan\\
$^{3}$ \quad NASA Goddard Space Flight Center, 8800 Greenbelt Rd, Greenbelt, MD 20771, USA\\
$^{4}$ \quad Max-Planck-Institut f$\rm \ddot{u}$r Kernphysik, Saupfercheckweg 1, D-69117 Heidelberg, Germany \\
$^{5}$ \quad Lawrence Livermore National Laboratory, 7000 East Avenue, Livermore, CA 94550, USA\\
$^{6}$ \quad Institute of Modern Physics, Chinese Academy of Sciences, Lanzhou 730000, China\\
$^{7}$ \quad University of Chinese Academy of Sciences, Beijing 100049, China
}

\corres{Correspondence: l.gu@sron.nl}

\abstract{Relevant uncertainties on theoretical atomic data are vital to determine the accuracy of plasma diagnostics in a number of areas including in particular the astrophysical study. We present a new calculation of the uncertainties on the present theoretical ion-impact charge exchange atomic data and X-ray spectra based on a set of comparisons with the existing laboratory data obtained in historical merged-beam, cold-target recoil-ion momentum spectroscopy, and electron beam ion traps experiments. The average systematic uncertainties are found to be $35-88$\% on the total cross sections, and $57-75$\% on the characteristic line ratios. The model deviation increases as the collision energy decreases. The errors on total cross sections further induce a significant uncertainty to the calculation of ionization balance for low temperature collisional plasmas. Substantial improvements of the atomic database and dedicated laboratory measurements are needed to get the current models ready for the X-ray spectra from the next X-ray spectroscopic mission. }


\begin{document}
%%%%%%%%%%%%%%%%%%%%%%%%%%%%%%%%%%%%%%%%%%

\section{Introduction}

Charge exchange plasma can be found in a broad range of astrophysical environments, including in particular the interfaces where the solar wind ions interact with neutrals in comets and planetary atmospheres \citep{lisse1996, crav1997, bode2007, br2007}, but potentially also in supernova remnants \citep{katsu2011, cumb2014}, star forming galaxies \citep{liu2010, zhang2014}, active galactic nuclei \citep{gu2017}, and clusters of galaxies \citep{gu2015, gu2018}. Modeling of the X-ray spectrum of charge exchange becomes possible recently by the efforts of \citet{smith2012} and \citet{2016A&A...588A..52G}. These models are crucial for interpreting the observations as well as to understand the physical sources that power the plasma. 

There is an increasing demand from the astronomical community that the plasma model should provide an estimate of the systematic uncertainties for the atomic data used. This is triggered by the accumulating evidence that the uncertainties from the atomic data, which are not accounted for at present, are as significant as the typical errors from instrumental calibration (see \citep{hitomiatomic} for a recent example). So far, there is no systematic estimate of the uncertainties of the existing charge exchange models, making it difficult to assess the accuracy of the scientific results obtained with these models. 

Most of the charge exchange reaction rates in existing models are obtained in theoretical calculations,  with only a few laboratory benchmarks performed by several groups with various experimental methods (see e.g., cross-beam/merged-beam neutral setups:~\citep{dijkkamp1985,bodowits2004,trassinelli2012}; tokamak and laser-produced plasmas:~\citep{rosmej2006,2015Beiersdorfer,2015Lepson}; cold-target recoil-ion momentum spectroscopy (COLTRIM):~\citep{2002Fischer,ali2005,ali2010,xue2014,ali2016}; electron beam ion trap (EBIT):~\cite{2000Beiersdorfer,2005Wargelin,2008Allen,2010Leutenegger,2014Martinez,2016Shah,2019DobrodeyPHD}). 
A recent comparison using the data from the EBIT measurements \citep{2018ApJ...868L..17B} showed that the model and the laboratory spectra differ significantly in both line energies and strengths, for the L-shell charge exchange between nickel ions and neutral particles. 
Another recent example is the COLTRIMS measurement by \citet{xu2021} showed that the model calculations might differ from the measurements by $20-50$\% for the state-selective cross sections of Ne$^{8+}$ and Ne$^{9+}$ charge exchange. 
In this work, we compile a sample of existing laboratory measurements on charge exchange total cross sections, state-selective cross sections, as well as characteristic X-ray line ratios, and put forward a systematic assessment of the model accuracy. 

This paper is arranged as follows. In Section 2 we describe the sample and the results of the benchmark, and in Section 3 we discuss the potential improvement with future EBIT and COLTRIMS measurements. The benchmark is directly applied to the charge exchange model and atomic data \citep{2016A&A...588A..52G} in the SPEX \citep{1996spex} software. Throughout the paper, the errors are given at a 68\% confidence level.

%%%%%%%%%%%%%%%%%%%%%%%%%%%%%%%%%%%%%%%%%%
\section{Methods and Results}

\subsection{Total cross sections}

\begin{figure}[H]	
\widefigure
\includegraphics[width=12.0 cm]{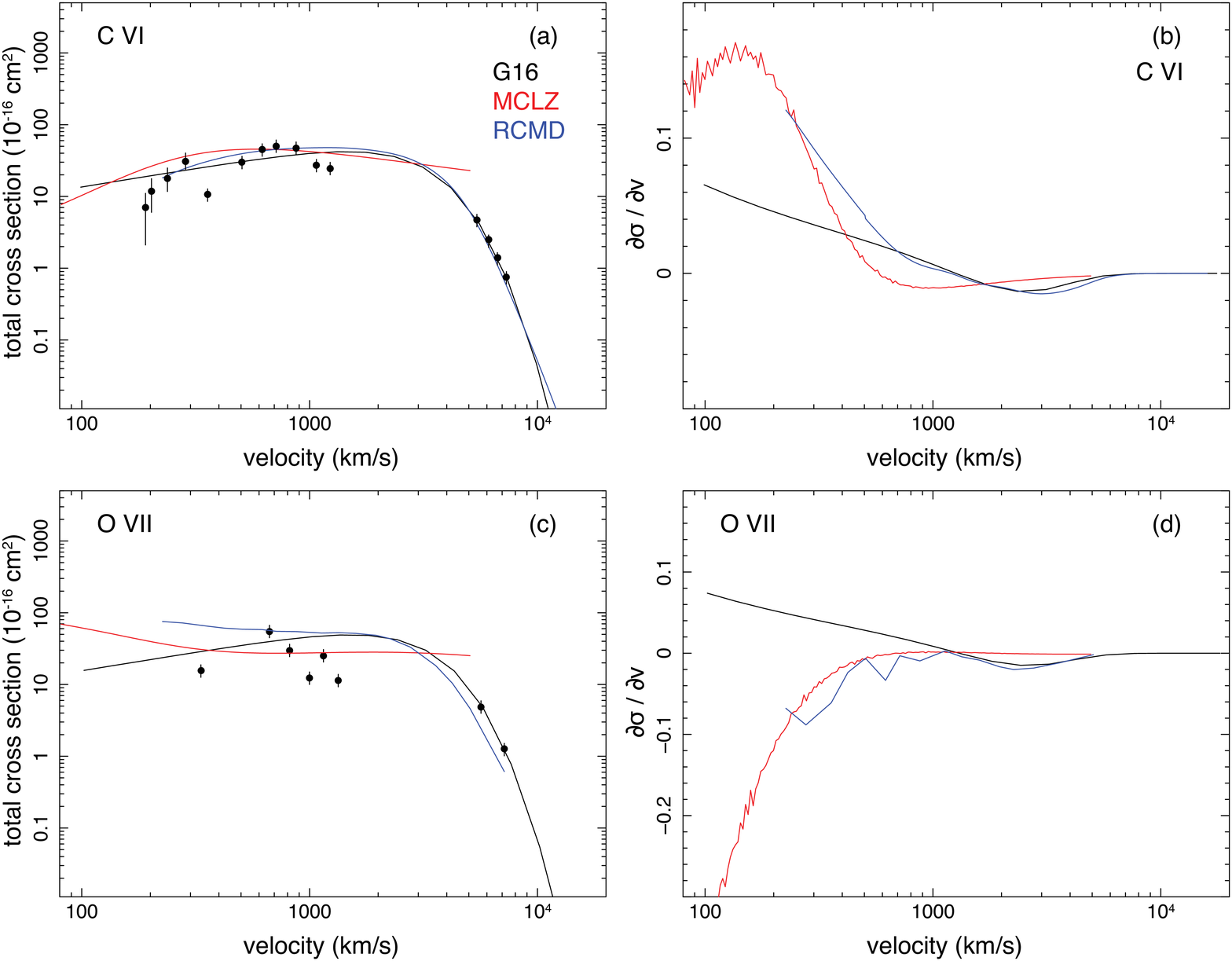}
\caption{Total cross sections as a function of collision velocity and the cross section derivatives with respect to the velocity 
for $\rm C^{6+}$ (a,b) and $\rm O^{7+}$ (c,d) ions interacting with hydrogen atoms, resulting in C VI and O VIII ions.
The data points are experimental results from \citet{1979JPhB...12.3763G, 1982PhRvA..26.1892P, 1983PhST....3..124P, 1979PhRvA..19..515M} . Approximate errors of 15\% \citep{1979PhRvA..19..515M} are shown, except for the low energy (< 500 km s$^{-1}$) data of C VI for which the actual errors were reported in the original paper. The solid lines are the model values from the calculations with the G16 (black), MCLZ (red), and RCMD (blue) methods. The abbreviations are explained in the text.
\label{cxdata}}
\end{figure}

\end{paracol}
\begin{paracol}{1} % added <<<<<<<<<<<<<<<<
   \small
    \setlength{\arrayrulewidth}{0.2mm}
    \renewcommand{\arraystretch}{1.2}   
\onecolumn
\begin{longtable}{ccccccc}
\caption{\label{exp_data}Experimental data sample} \\
\hline
Reference &  type$^{a}$  & ion & theory data  \\
\hline
\endfirsthead
\centering
Reference &  type$^{a}$  & ion & theory data  \\
\hline
\endhead
%\tablefoottext{a}{Emissivities are in units of $10^{-23}$~ph\,m$^3$\,s$^{-1}$.}
\endfoot
\citet{exp1} & total & $\rm Li^{q+}$ (q=1-3) & G16 \\
\citet{exp2} & total & $\rm Li^{q+}$ (q=2-3), $\rm N^{q+}$ (q=2-5), $\rm Ne^{q+}$ (q=3-5) & G16 \\
\citet{1979JPhB...12.3763G} & total & $\rm B^{q+}$ (q=1-5), $\rm C^{q+}$ (q=1-4) & G16 \\
\citet{1979JPhB...12.3763G} & total & $\rm C^{q+}$ (q=5,6), $\rm N^{7+}$ & G16, MCLZ, RCMD \\
\citet{exp3} & total & $\rm B^{2+}$, $\rm C^{+}$, $\rm N^{+}$, $\rm Mg^{2+}$ & G16 \\
\citet{exp4} & total & $\rm B^{q+}$ (q=2-5), $\rm C^{q+}$ (q=3,4), $\rm N^{q+}$ (q=3,4), $\rm O^{q+}$ (q=5,6) & G16 \\
\citet{exp5} & total & $\rm B^{q+}$ (q=2-4), $\rm C^{q+}$ (q=2-4), $\rm N^{q+}$ (q=2-5), $\rm O^{q+}$ (q=2-5) & G16 \\
\citet{exp6} & total & $\rm C^{q+}$ (q=1-4), $\rm N^{q+}$ (q=1-5), $\rm O^{q+}$ (q=1-5), $\rm Si^{q+}$ (q=2-7) & G16 \\
\citet{exp7} & total & $\rm C^{2+}$ & G16 \\
\citet{1982PhRvA..26.1892P} & total & $\rm C^{q+}$ (q=3,4), $\rm O^{q+}$ (q=2-6) & G16 \\
\citet{1982PhRvA..26.1892P} & total & $\rm C^{q+}$ (q=5,6)& G16, MCLZ, RCMD \\
\citet{exp8} & total & $\rm C^{3+}$ & G16 \\
\citet{exp9} & total,nl & $\rm C^{q+}$ (q=3,4), $\rm N^{5+}$, $\rm O^{6+}$ & G16 \\
\citet{exp10} & total,nl & $\rm C^{3+}$ & G16 \\
\citet{1983PhST....3..124P} & total & $\rm C^{4+}$, $\rm N^{5+}$, $\rm O^{6+}$, $\rm Ne^{8+}$ & G16 \\
\citet{1983PhST....3..124P} & total & $\rm C^{q+}$ (q=5,6), $\rm N^{q+}$ (q=6,7), $\rm O^{q+}$ (q=7,8), $\rm Ne^{q+}$ (q=9,10) & G16, MCLZ, RCMD \\
\citet{exp11} & total,nl & $\rm C^{q+}$ (q=3,4), $\rm N^{5+}$, $\rm O^{6+}$ & G16 \\
\citet{exp12} & total,nl & $\rm C^{4+}$ & G16 \\
\citet{exp13} & total,nl & $\rm C^{4+}$ & G16 \\
\citet{exp14} & total & $\rm N^{+}$, $\rm O^{+}$ & G16 \\
\citet{exp15} & total & $\rm O^{+}$ & G16 \\
\citet{1979PhRvA..19..515M} & total & $\rm B^{q+}$ (q=2-5), $\rm C^{q+}$ (q=3,4), $\rm N^{q+}$ (q=3,4) & G16 \\
\citet{1979PhRvA..19..515M} & total & $\rm O^{q+}$ (q=3-6), $\rm Si^{q+}$ (q=4-9), $\rm Fe^{q+}$ (q=4-15) & G16 \\
\citet{1979PhRvA..19..515M} & total & $\rm O^{q+}$ (q=7,8) & G16, MCLZ, RCMD \\
\citet{exp16} & total & $\rm O^{5+}$ & G16 \\
\citet{exp17} & total & $\rm Ne^{q+}$ (q=2-4), $\rm Ar^{q+}$ (q=2-4,6) & G16 \\
\citet{exp18} & total & $\rm Si^{q+}$ (q=2-7) & G16 \\
\citet{exp19} & nl & $\rm O^{3+}$ & G16 \\
\citet{exp20} & total & $\rm Ne^{3+}$ & G16 \\
\citet{exp21} & total & $\rm Ne^{4+}$ & G16 \\
\citet{exp22} & total & $\rm Si^{3+}$ & G16 \\
\citet{exp23} & total & $\rm C^{3+}$ & G16 \\
\citet{exp24} & total & $\rm Ne^{2+}$ & G16 \\
\citet{exp25} & total & $\rm B^{4+}$ & G16 \\
\citet{exp26} & total & $\rm N^{4+}$ & G16 \\
\hline
\end{longtable}
\begin{itemize}
\item[$a$] total = total cross section, nl = $nl$-resolved cross section
\end{itemize}

\end{paracol}% added <<<<<<<<<<<<<<<<

\begin{paracol}{2} % added <<<<<<<<<<<<<<<<
    \switchcolumn % added <<<<<<<<<<<<<<<<

First we compare the SPEX calculations with the existing laboratory results for a number of ions on their total cross sections for atomic hydrogen targets. The SPEX atomic data is not one uniform set of theoretical calculations, but a mixture of three different types of approaches: (1) the rates derived with the empirical scaling reported in \citet{2016A&A...588A..52G} (G16 hereafter), which was based on a numerical approximation to a collection of historical theoretical and experimental rates; (2) the multi-channel Landau-Zener method (hereafter MCLZ) reported in \citet{2016ApJS..224...31M}. The atomic data generated by MCLZ are also publicly available in the Kronos database \footnote{https://www.physast.uga.edu/research/stancil-group/atomic-molecular-databases/kronos}; and (3) the recommended values (hereafter RCMD) based on dedicated calculations, including in most cases the quantum-mechanical and classical molecular-orbital close-coupling methods, and the atomic-orbital close-coupling method. The G16 approach can calculate for any ions with a given atomic number and charge, the MCLZ data cover most of the H- and He- like ions with atomic number up to 30, and the RCMD rates are available for a small set of key ions, e.g., O VII \citep{2012JPhB...45w5201W}, N VII \citep{2011PhRvA..84b2711W}, and C VI \citep{2012JPhB...45x5202N}.

All the three datasets are tested when the corresponding theoretical cross sections ( $\sigma_{\rm theo}$) and experimental cross sections ($\sigma_{\rm exp}$) are available. Examples are shown in Figure~\ref{cxdata} for the C VI and O VII data. For C VI, the three calculations converge at the energy range from $\sim 100$ eV/amu to $4\times10^4$ eV/amu, while the MCLZ data do not cover higher energies and the G16 and RCMD data miss the low energy part. For O VII, the difference between the three calculations becomes more significant than in the case of C VI. The cross section derivatives shown in Figure~\ref{cxdata} indicate that the differences in the shapes of the three theoretical calculations become in general larger at lower collision energies.

\begin{figure}[H]	
\widefigure
\includegraphics[width=12.0 cm]{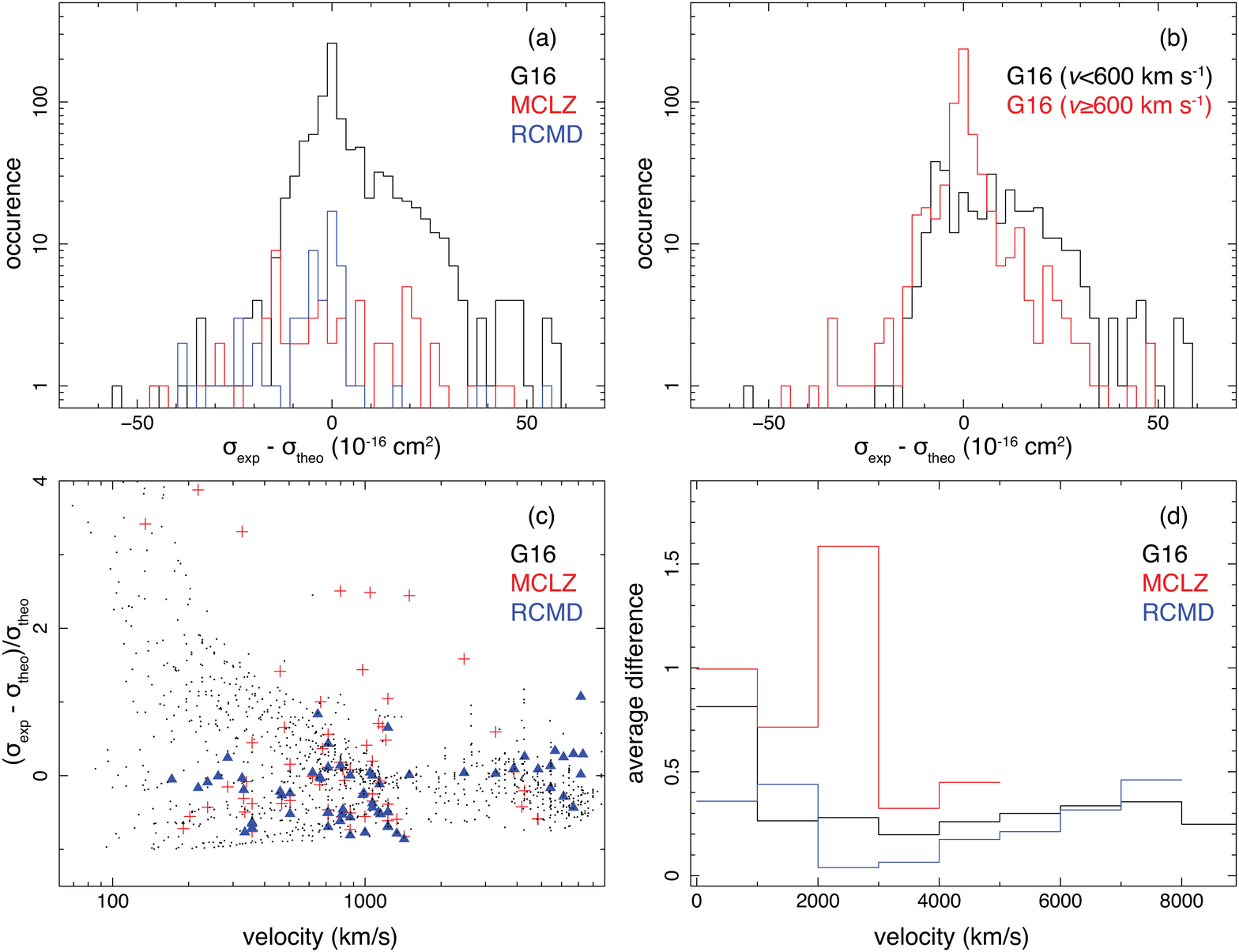}
\caption{Distributions of the absolute (upper) and relative (lower) deviations 
of the theoretical cross sections from the experimental results obtained with the measurements summarized in Table~\ref{exp_data}. (a) Diagrams of the absolute errors for the G16 (black), MCLZ (red), and RCMD (blue) theories. (b) Diagrams of the absolute errors for G16 for low collision velocities (black) and high velocities (red).
(c) The relative deviations for the G16 (black points), MCLZ (red crosses), and RCMD (blue triangles) 
calculations. (b) The average deviations in absolute values for the three methods in each velocity interval.  
\label{cxsum}}
\end{figure}

In Figure~\ref{cxsum}, we plot the distributions of the absolute errors $\sigma_{\rm exp}$ - $\sigma_{\rm theo}$ of the theoretical models. The standard deviations of the absolute errors are 1.4 $\times 10^{-15}$ cm$^{-2}$, 2.0 $\times 10^{-15}$ cm$^{-2}$, and 1.3 $\times 10^{-15}$ cm$^{-2}$, for the G16, MCLZ, and RCMD calculations, respectively. As shown in Figure~\ref{cxsum}(b), the absolute errors of G16 become more scattered, and on average larger, at lower collision velocities. The standard deviations of the error distributions are 1.8 $\times 10^{-15}$ cm$^{-2}$ for $v < 600$ km s$^{-1}$, and 0.9 $\times 10^{-15}$ cm$^{-2}$ for $v \geq 600$ km s$^{-1}$. 

We also summarize the relative deviations ($\sigma_{\rm exp}$ - $\sigma_{\rm theo}$) / $\sigma_{\rm theo}$ of the three calculations in Figure~\ref{cxsum}. The average absolute values of the fractional deviations are 55\%, 88\%, and 35\% for the G16, MCLZ, and RCMD datasets. Similar to G16, the MCLZ calculation also has larger relative errors for low velocity collisions, while the RCMD calculation shows fairly constant deviations for the velocity range considered. For high-energy collisions of $v > 3000$ km s$^{-1}$, the three methods show reasonable agreement with the laboratory results within uncertainties $< 50$\%.

%The deviation appears to be slightly velocity dependent for in particular the G16 and MCLZ results. The cross sections calculated with G16 for low velocity collisions ($v < 1000$ km s$^{-1}$) are nearly twice more biased than those for higher velocities. This is probably caused by the fact that G16 tends to underestimate the cross sections for the very low velocity range. A similar case might also hold for the MCLZ calculations though the data points on the MCLZ comparison are much more sparse. The RCMD dataset shows a relatively constant average deviation for the velocity range considered. For high-energy collisions of $v > 3000$ km s$^{-1}$, the three methods are consistent with the laboratory results within uncertainties $< 50$\%. 

The laboratory results should have their own uncertainties, however, these values are available for only a part of the measurements. Here we provide a rough estimate of the combined measurement uncertainty. The mean systematic uncertainties on the cross sections measured in, e.g., \citet{1979PhRvA..19..515M, dra2011, cab2020}, are approximately 15\% for the energy range considered. Assuming that this value can be applied to the other laboratory results, the measurement uncertainties
are about 1\% for the sample used in testing the G16 calculation, and $\sim 4$\% for the MCLZ and RCMD results. These relatively minor uncertainties can be accepted as the errors of the theoretical deviations obtained above (e.g., 55\%, 88\%, and 35\% for the G16, MCLZ, and RCMD approaches).

\begin{figure}[H]	
\widefigure
\includegraphics[width=12.0 cm]{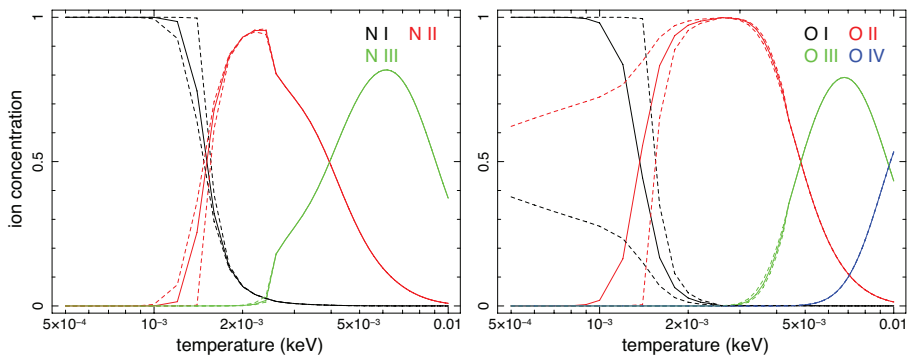}
\caption{Charge state distributions of N (left) and O (right) as a function of equilibrium temperature for the CIE plasma, calculated with SPEX version 3.06.01. The dashed lines show the calculations when the charge exchange recombination rates are changed by 50\%, while the other ionization and recombination data are kept intact.   
\label{cxcon}}
\end{figure}

The total charge exchange cross section is needed not only for calculating the charge exchange emission, but also to derive the ionization concentration for general cosmic plasmas in collisional ionization or photoionization equilibrium. The uncertainties in the theoretical calculation would introduce systematic uncertainties to the charge state distribution for the low temperature plasmas where ions and neutral atoms coexist. As shown in Figure~\ref{cxcon}, we present two test cases on the concentration calculations of N and O ions in collisional ionization equilibrium (CIE). Here we assume an uncertainties of 50\% on the charge exchange recombination rates. The induced errors on the charge distributions of N I and O I would become 10\% and 60\% at an equilibrium temperature of 1.2 eV. The difference between N I and O I errors reflects the different relative contribution of charge exchange to the total recombination in the concentration calculation. This result suggests that the charge exchange atomic data is vital to the modeling accuracy of lowly ionized species for collisional plasmas. It is expected that similar uncertainties would also apply to the photoionization modeling which includes the charge exchange component in the same way.

\subsection{Cross sections for the peak $nl$ shells}

\begin{figure}[H]	
\widefigure
\includegraphics[width=12.0 cm]{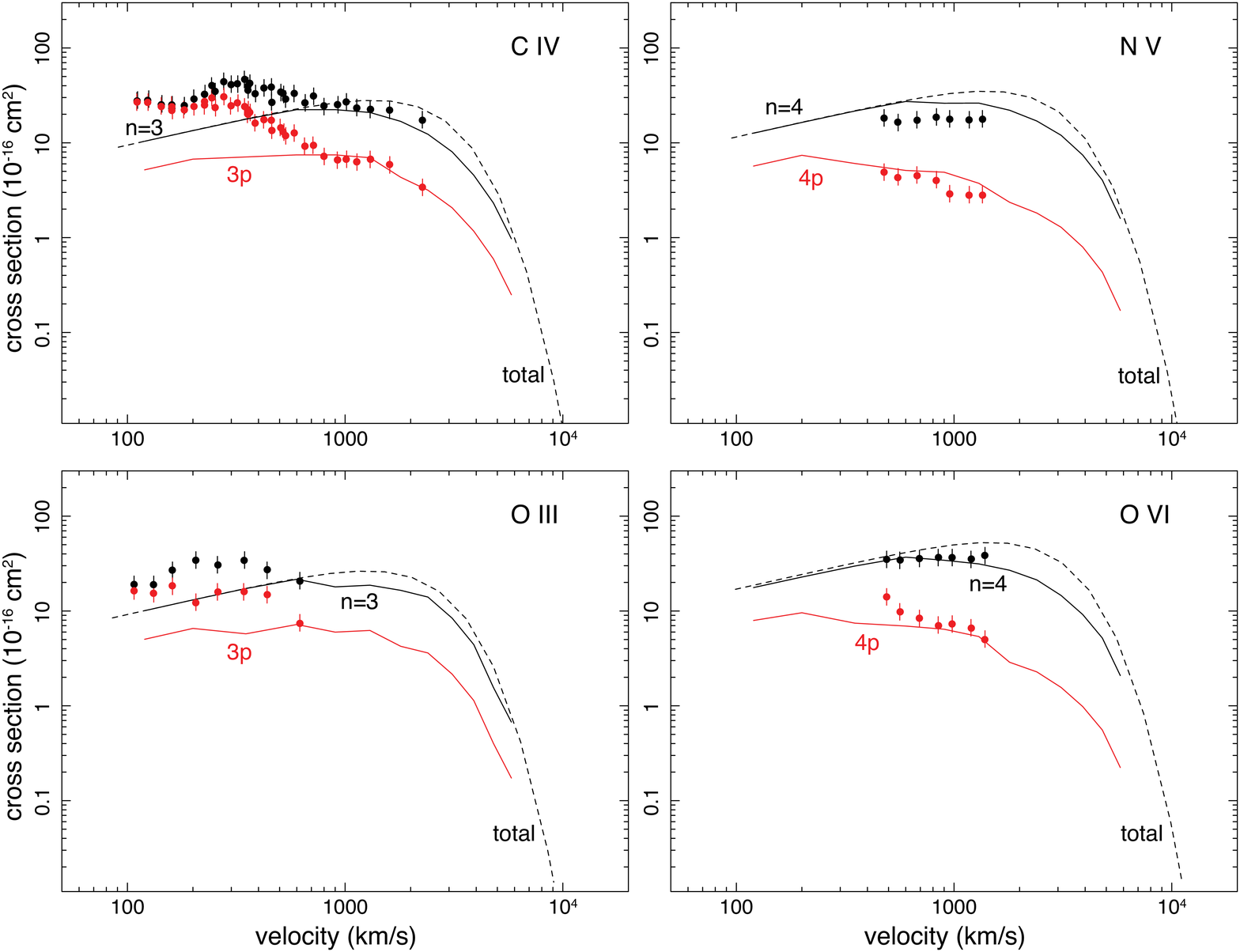}
\caption{State-selective cross sections as a function of collision velocity for C IV, N V, O III, and O VI. The data points are taken from merged-beam experiments (see Table~\ref{exp_data} for detail), for the peak $n$ shells (black) and the $np$ subshells (red). Approximate errors of 15\% \citep{1979PhRvA..19..515M} are shown. The black solid lines are the G16 calculations of the peak $n$ shells, and the red lines are the G16 data for the $np$ subshells. The dash lines are the G16 calculations of the total cross sections. 
\label{nl}}
\end{figure}  

Next we examine the state-resolved cross sections. The selective population of high-$n$ levels of the recombining ions is known to be a characteristic property of the charge exchange reaction. The distribution functions on the quantum numbers $n$ and $l$ are key to the calculation of the spectrum, though the present theory still cannot fully reproduce the $nl$ distributions measured in the laboratory \citep{hr1, 2000PhRvL..85.5090B, 2018ApJ...868L..17B}. 

As shown in Figure~\ref{nl}, we compare the laboratory measurements of four reactions with theoretical calculations using the G16 method. G16 is the only calculation available in SPEX for the ions tested. It defines empirically $n$ of the most populated levels as functions of the collision velocity, charge, and ionization potential. For the four test cases, G16 successfully predicts the peak $n$: $n=3$ for C IV and O III, $n=4$ for N V and O VI. The cross sections of the peak $n$ levels, however, show deviations from the G16 values at the low energies. For C IV and O III, the measured values for $v = 100$ km s$^{-1}$ are higher by a factor of $\sim 2.5$ than the theoretical ones. This is probably because that the G16 method underestimates the total cross sections at low energies as already shown in Figure~\ref{cxsum}. For $v > 500$ km s$^{-1}$, the G16 calculations become consistent with the measurements within 40\% for the peak $n$.

To assess the $l-$distribution function, in Figure~\ref{nl} we also compare the cross sections of the $np$ subshells. The $l$-distribution defined in G16 is a smooth function that switches as a function of velocity between the different empirical $l$ distributions introduced in \citet{1985PhR...117..265J} (see also Eqs.4-8 and Appendix B in \citep{2016A&A...588A..52G}). The G16 cross sections on the $np$ shells are lower, by a factor of 2-5, than the experimental values for $v < 500$ km s$^{-1}$. The deviations become again much smaller at higher velocities. To summarize above, the G16 method could reproduce the $nl$-resolved cross sections for the test cases with an accuracy of $\sim 40$\% for $v > 500$ km s$^{-1}$, while for the low velocity collision, the G16 cross sections, as well as the line intensities calculated based on the atomic data, are much less reliable.

\subsection{Line ratios}

\begin{figure}[H]	
\widefigure
\includegraphics[width=12.0 cm]{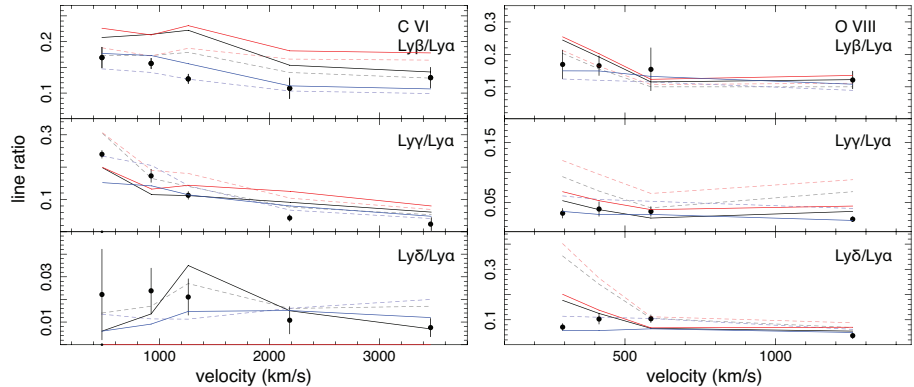}
\caption{Comparison of experimental and theoretical line ratios for the C$^{6+}$ (left) and O$^{8+}$ (right) reactions. The experimental data from \citet{hr3} and \citet{hr2} are plotted as data points, and the predictions from G16, MCLZ, and RCMD are shown in black, red, and blue curves. 
The dashed curves show the calculations of C$^{6+}$ and O$^{8+}$ collisions with H atoms, and the solid lines show the collisions with Kr atoms obtained using the scaling of \citet{leung2018}. 
\label{lineratio}}
\end{figure}  

The large ratios between $1s-np$ ($n>2$) and $1s-2p$ lines are often used as a characteristic diagnostics of the highly-charged charge exchange plasma \citep{gu2015,gu2018}. It is known that the line ratios would decrease with the increasing collision velocity, because a high speed collision might yield captures on high angular momentum states, producing more $1s-2p$ transitions through cascade. So the line ratios can often be utilized as a probe of collision velocity \citep{bei2000, otra2006}. The accuracy of the velocity measurement is therefore determined by the quality of the atomic data.

In Figure~\ref{lineratio} we plot the comparison of the line ratio calculations and experiments for C VI and O VIII. The experimental data are taken from the beam-gas measurements by \citet{hr3} for C VI and \citet{hr2} for O VIII. A caveat of the comparison is that these experiments used Kr atom as a target while the original theoretical calculations are based on capture from H atom. As reported in \citet{leung2018}, the line ratios from Kr and H collisions are somewhat different in particular for the low-energy regime, even though the ionization potentials of Kr and H atoms are nearly the same. To compensate this discrepancy, we calculate the H-to-Kr scalings as a function of velocities on both C VI and O VIII line ratios using the theoretical results reported in \citet{leung2018} (in their Figures 3 and 6), and apply the scalings to the G16, MCLZ, and RCMD line ratios. The scaled line ratios should represent a better approximation to the collisions with the Kr target.

As shown in Figure~\ref{lineratio}, the experiments and calculations yield the same peak $n$, $n=4$ for C VI and $n=5$ for O VIII, though the line ratios still differ at several velocities. One of the main discrepancies occurs between the scaled G16/MCLZ and the lab data for the C VI Ly$\beta$/Ly$\alpha$ line ratio, where the two theoretical values exceed the measured one by about 70\% at $v=1000$ km s$^{-1}$. The RCMD calculation shows better agreement with the lab values on this line ratio.

\end{paracol}
\begin{paracol}{1} % added <<<<<<<<<<<<<<<<

\onecolumn
\begin{longtable}[h]{l@{\hskip 0.1in}c@{\hskip 0.1in}c@{\hskip 0.1in}c@{\hskip 0.1in}c@{\hskip 0.1in}c@{\hskip 0.1in}c@{\hskip 0.1in}c} 
\caption{\label{lines_ratio}Experimental and theoretical Line ratios} \\
\hline \hline
ion & $v$ (km s$^{-1}$) & ratio & experiment   & G16  & MCLZ  & RCMD & reference \\
\hline 
\endfirsthead
\centering
ion & $v$ (km s$^{-1}$) & ratio & experiment   & G16  & MCLZ  & RCMD & reference$^{a}$ \\
\hline
\endhead
\hline
%\tablefoottext{a}{Emissivities are in units of $10^{-23}$~ph\,m$^3$\,s$^{-1}$.}
\endfoot
N VII    & 794      & Ly$\beta$/Ly$\alpha$ & 0.76   & 0.12 & 0.13  & 0.10 & 1 \\
         &          & (Ly$\gamma$+Ly$\delta$)/Ly$\alpha$ & 0.62 & 0.18 & 0.20 & 0.29 & \\
O VII    & 724      & Ly$\beta$/Ly$\alpha$ & 0.19   & 0.07 & 0.09  & 0.07 & \\
         &          & (Ly$\gamma$+Ly$\delta$)/Ly$\alpha$ & 0.24 & 0.11 & 0.47 & 0.07 & \\
O VIII   & 774      & Ly$\beta$/Ly$\alpha$ & 0.13   & 0.11 & 0.10  & 0.11 & \\
         &          & (Ly$\gamma$+Ly$\delta$)/Ly$\alpha$ & 0.17 & 0.15 & 0.18 & 0.14 & \\
Ne IX    & 743      & Ly$\beta$/Ly$\alpha$ & 0.04   & 0.04 & 0.12  & $-$  & \\
         &          & (Ly$\gamma$+Ly$\delta$)/Ly$\alpha$ & 0.05 & 0.05 & 0.18 & $-$  & \\
Ne X     & 783      & Ly$\beta$/Ly$\alpha$ & 0.12   & 0.08 & 0.08 & 0.08  & \\
         &          & (Ly$\gamma$+Ly$\delta$)/Ly$\alpha$ & 0.11 & 0.06 & 0.08 & 0.04 & \\
O VIII   & 293      & Ly$\beta$/Ly$\alpha$ & 0.169$\pm$0.044 & 0.244$^{b}$ &	0.254 &	0.149   & 2 \\
         &          & Ly$\gamma$/Ly$\alpha$& 0.032$\pm$0.008 & 0.053 &	0.068 &	0.035 & \\
         &          & Ly$\delta$/Ly$\alpha$& 0.071$\pm$0.014 & 0.177 &	0.201 &	0.057 & \\
         &          & Ly$\epsilon$/Ly$\alpha$& 0.0065$\pm$0.003 & 0.054 & 0.0061 & 0.027 & \\
O VIII   & 414      & Ly$\beta$/Ly$\alpha$ & 0.165$\pm$0.030 & 0.192 &	0.202 &	0.149   & \\
         &          & Ly$\gamma$/Ly$\alpha$& 0.039$\pm$0.012 & 0.038 &	0.053 &	0.030 & \\
         &          & Ly$\delta$/Ly$\alpha$& 0.103$\pm$0.02 & 0.125 &	0.138 &	0.057 & \\
         &          & Ly$\epsilon$/Ly$\alpha$& 0.005$\pm$0.0076 & 0.031 & 0.0024 & 0.019 & \\
O VIII   & 586      & Ly$\beta$/Ly$\alpha$ & 0.154$\pm$0.006 & 0.115 &	0.123 &	0.132   & \\
         &          & Ly$\gamma$/Ly$\alpha$& 0.035$\pm$0.008 & 0.024 &	0.038 &	0.030 & \\
         &          & Ly$\delta$/Ly$\alpha$& 0.104$\pm$0.015 & 0.066 &	0.068 &	0.064 & \\
         &          & Ly$\epsilon$/Ly$\alpha$& 0.0048$\pm$0.0061 & 0.015 & 0.00086 & 0.014 & \\
O VIII   & 1256      & Ly$\beta$/Ly$\alpha$ & 0.121$\pm$0.027 & 0.122 &	0.135 &	0.108   & \\
         &          & Ly$\gamma$/Ly$\alpha$& 0.022$\pm$0.004 & 0.035 &	0.044 &	0.020 & \\
         &          & Ly$\delta$/Ly$\alpha$& 0.037$\pm$0.011 & 0.055 &	0.071 &	0.050 & \\
         &          & Ly$\epsilon$/Ly$\alpha$& 0.0048$\pm$0.0028 & 0.023 & 0.00045 & 0.0090 & \\
C VI     & 477      & Ly$\beta$/Ly$\alpha$ & 0.169$\pm$0.023 & 0.208$^{b}$ &	0.226 &	0.177   & 3 \\
         &          & Ly$\gamma$/Ly$\alpha$& 0.240$\pm$0.012 & 0.198 &	0.199 &	0.152 & \\
         &          & Ly$\delta$/Ly$\alpha$& 0.022$\pm$0.020 & 0.0062 &	2.8e-6 &	0.0061 & \\
C VI     & 924      & Ly$\beta$/Ly$\alpha$ & 0.157$\pm$0.012 & 0.214 &	0.213 &	0.173   &  \\
         &          & Ly$\gamma$/Ly$\alpha$& 0.173$\pm$0.023 & 0.115 &	0.132 &	0.142 & \\
         &          & Ly$\delta$/Ly$\alpha$& 0.024$\pm$0.009 & 0.014 &	2.7e-6	& 0.0091 & \\
C VI     & 1262      & Ly$\beta$/Ly$\alpha$ & 0.128$\pm$0.009 & 0.222 &	0.231 &	0.157   &  \\
         &          & Ly$\gamma$/Ly$\alpha$& 0.113$\pm$0.012 & 0.112 &	0.144 &	0.114 & \\
         &          & Ly$\delta$/Ly$\alpha$& 0.021$\pm$0.008 & 0.035 &	2.2e-6 &	0.015 & \\
C VI     & 2185      & Ly$\beta$/Ly$\alpha$ & 0.109$\pm$0.019 & 0.154	&	0.182 &	0.114  &  \\
         &          & Ly$\gamma$/Ly$\alpha$& 0.043$\pm$0.011 & 0.091	& 0.125	&	0.080 & \\
         &          & Ly$\delta$/Ly$\alpha$& 0.011$\pm$0.006 & 0.015	& 6.7e-7 &	0.015 & \\      
C VI     & 3466      & Ly$\beta$/Ly$\alpha$ & 0.130$\pm$0.021 & 0.141	&	0.178 &	0.108&  \\
         &          & Ly$\gamma$/Ly$\alpha$& 0.024$\pm$0.018 & 0.061 &	0.080 &		0.048 & \\
         &          & Ly$\delta$/Ly$\alpha$& 0.0076$\pm$0.004 & 0.007 &	5.0e-7	&	0.012 & \\              
O VII    & low      & He$_{\rm high}$/He$\alpha$ & 0.167     & 0.168 & 0.152 & 0.058 & 4 \\
Ne IX    & low      & He$_{\rm high}$/He$\alpha$ & 0.162     & 0.161 & 0.133 & $-$  & \\
Ar XVII  & low      & He$_{\rm high}$/He$\alpha$ & 0.191     & 0.133 & $-$ & $-$ & \\
Fe XXV   & low      & He$_{\rm high}$/He$\alpha$ & 0.267     & 0.156 & 0.079 & $-$ & \\
O VIII   & low      & Ly$_{\rm high}$/Ly$\alpha$ & 1.006     & 0.786 & 0.887 & 0.366 & \\
Ne X     & low      & Ly$_{\rm high}$/Ly$\alpha$ & 1.207     & 0.690 & 0.865 & 0.210 & \\
Mg XII   & low      & Ly$\beta$/Ly$\alpha$ & 0.227$\pm$0.040 & 0.179 & 0.205 & $-$ & 5 \\ 
Mg XII   & low      & Ly$\gamma$/Ly$\alpha$ & 0.133$\pm$0.022 & 0.070 & 0.083 & $-$ & \\
Mg XII   & low      & Ly$\delta$/Ly$\alpha$ & 0.044$\pm$0.015 & 0.038 & 0.046 & $-$ & \\
Mg XII   & low      & Ly$\epsilon$/Ly$\alpha$ & 0.095$\pm$0.015 & 0.028 & 0.030 & $-$ & \\
Mg XII   & low      & Ly$\zeta$/Ly$\alpha$ & 0.030$\pm$0.018 & 0.221 & 0.120 & $-$ & \\
Mg XII   & low      & Ly$\eta$/Ly$\alpha$ & 0.080$\pm$0.014 & 0.091 & 0.287 & $-$ & \\
S XVI    & low      & Ly$\beta$/Ly$\alpha$ & 0.203$\pm$0.070 & 0.153 & 0.171 & $-$ & 5 \\ 
S XVI    & low      & Ly$\gamma$/Ly$\alpha$ & 0.082$\pm$0.016 & 0.055 & 0.064 & $-$ & \\ 
S XVI    & low      & Ly$\delta$/Ly$\alpha$ & 0.053$\pm$0.011 & 0.028 & 0.033 & $-$ & \\ 
S XVI    & low      & Ly$\epsilon$/Ly$\alpha$ & 0.053$\pm$0.008 & 0.017 & 0.020 & $-$ & \\ 
S XVI    & low      & Ly$\zeta$/Ly$\alpha$ & 0.016$\pm$0.005 & 0.012 & 0.014 & $-$ & \\ 
S XVI    & low      & Ly$\eta$/Ly$\alpha$ & 0.029$\pm$0.008 & 0.024 & 0.014 & $-$ & \\ 
S XVI    & low      & Ly$\theta$/Ly$\alpha$ & 0.111$\pm$0.019 & 0.149 & 0.101 & $-$ & \\ 
S XVI    & low      & Ly$\iota$/Ly$\alpha$ & 0.165$\pm$0.024 & 0.058 & 0.165 & $-$ & \\ 
\hline
\hline
\end{longtable}
\begin{itemize}
\item[$a$] references 1 = \citet{hr1}; 2 = \citet{hr2}: 3 = \citet{hr3}; 4 = \citet{hr4}; 5 = \citet{hr5}.
\item[$b$] H-to-Kr scaling has been applied to the theoretical line ratios for O VIII and C VI; see text for details.
\end{itemize}

\end{paracol}% added <<<<<<<<<<<<<<<<

\begin{paracol}{2} % added <<<<<<<<<<<<<<<<
    \switchcolumn % added <<<<<<<<<<<<<<<<

A more extensive comparison can be seen in Table~\ref{lines_ratio}. It is a compilation of several laboratory efforts including the recent electron beam ion trap devices with X-ray spectral analysis carried out at both low and high resolutions. The EBIT devices simulate charge exchange reactions only at low collision energies. The average relative discrepancies (experiment - theory / theory) of the line ratios are 0.63, 0.77, and 0.54 for the G16, MCLZ, and RCMD calculations, respectively. For the peak $n$ shell, the average discrepancies are 0.61, 0.81, and 0.56 for the three models. These differences are significantly larger than those on the modeling of collisional ionization equilibrium plasmas ($\sim 10-40$\%, \citep{hitomiatomic, g19, g20}), suggesting that the state-of-the-art charge exchange spectral models, even with dedicated theoretical calculations, are still less reliable than those for the CIE plasma.

%%%%%%%%%%%%%%%%%%%%%%%%%%%%%%%%%%%%%%%%%%
\section{Discussion and ending remarks}

Based on a large sample of laboratory measurements, we have systematically compared the commonly used charge exchange atomic data to the experimental results. The G16, MCLZ,
and RCMD calculations utilized in the SPEX code do not fully reproduce the measurements, with 
notable, and likely velocity-dependent discrepancies in both total cross sections, state-resolved cross sections, and line ratios in the X-ray spectra. While the ease of use
of the present CX model is beneficial for the X-ray astronomical community, it should be 
used with cautions in particular for non-charge-exchange experts. The unresolvable 
disagreements call for advanced theoretical calculations for especially the low collision energy regime, in combination with more laboratory measurements with in particular the EBIT and COLTRIMS facilities. 

%{\bf Chintan: could you please put some text on prospect EBIT on CX }

The previous EBIT experiments have provided relevant benchmark to the predicted cross sections for electron capture into specific principal quantum number states $n$. However, a comparison with the angular-momentum $l$-resolved cross sections is challenging, as they depend on the collision energy; and the EBIT measurements are limited to low collision energies (< 10 eV/u)~\citep{2000Beiersdorfer}. 
Besides, the charge exchange process not only produces X-ray lines but also generates lines in the ultraviolet and optical band as the Rydberg levels populated by charge exchange relax through radiative cascades to the ground state of the ion. 
Thus, the simultaneous measurements of EUV and optical charge exchange cascade photons at the EBIT would be of interest, and they could provide additional information on the population of $nl$-states for the plasma modeling~\citep{2019DobrodeyPHD}.
Furthermore, possible multi-electron capture contribution from the molecular targets used in the EBIT measurements can also be avoided by using an atomic hydrogen target, where only single-electron capture can occur~\citep{2016leute}. Atomic hydrogen is of particular interest as it is also a most abundant neutral element in the Universe, and it makes a comparison between laboratory measurements and astrophysical observations more reliable. 

%{\bf Ruitian: could you please put some text on prospect COLTRIMS on CX }

Besides EBIT, The COLTRIMS and beam-gas experiments have been providing reliable measurements on velocity-dependent total and state-resolved cross sections. The improvement in the momentum measurement technique allows $nl$ selectivity, and for a few cases, it might even be able to resolve the spin state. The state-of-the-art measurement accuracy is about 11\% for both the total and $nl$-resolved cross sections \citep{2021han}. %Further improvements are soon expected. 

A systematic measurement of the cosmic abundant ions with the COLTRIMS facilities, in combination with simultaneous EBIT X-ray spectroscopy, is desirable for the astronomical community.
A consistent and continuous effort will be needed to ensure that the charge exchange atomic data will be ready for the high resolution X-ray spectra taken with next generation missions, XRISM (launch due in 2023, \citep{xrism}) and Athena (early 2030s, \citep{athena}).

%At Institute of Modern Physics, Chinese Academy of Science, employing the newly-built high-resolution reaction microscope at low energy accelerators, we are planning to perform series measurements of charge exchange processes with $nl$- resolved state selectivity (even with spin state resolved in some cases) for highly charged ions impact on neutral atoms or molecules in astrophysical environments. These measurements can provide reliable $nl$-resolved cross sections. Moreover, with these data the charge exchange X-ray emission spectra and line ratios could be obtained by considering radiative cascades. These are very helpful for modelling astrophysical charge exchange plasma.

Assessing uncertainties carried by the theoretical atomic data is also vital to the success of the upcoming missions. The atomic physics and plasma code community has already begun this work, with a persistent effort on the evaluation of the errors on electron impact excitation and transition probability data \citep{bau2013, loch2013, hitomiatomic, yu2018, foster2020, morisset2020}, as well as errors on photon impact data and modeling \citep{med2016}. One implication from the aforementioned works, including the present work on the charge exchange modeling, is that the classical assumption of a constant model uncertainty (e.g., 20\% on line emissivity) is no longer valid, since the uncertainties are proven to vary significantly with the underlying model and its key parameters. 

\section*{Acknowledgements}
SRON is supported financially by NWO, the Netherlands
Organization for Scientific Research. C.S. acknowledge support from an appointment to the NASA Postdoctoral Program at the NASA Goddard Space Flight Center, administered by the Universities Space Research Association, under contract with NASA, by the Lawrence Livermore National Laboratory (LLNL) Visiting Scientist and Professional Program Agreement, and by Max-Planck-Gesellschaft (MPG). 
The research leading to these results has received
funding from the European Union’s Horizon 2020 Programme under the
AHEAD2020 project (grant agreement n. 871158).

%%%%%%%%%%%%%%%%%%%%%%%%%%%%%%%%%%%%%%%%%%
\end{paracol}
\reftitle{References}

\end{document}